\documentclass[conference]{IEEEtran}
\usepackage{stmaryrd}
\usepackage{amssymb}
\usepackage{mathrsfs}
\usepackage{amsfonts}  
\usepackage[dvips,final]{graphicx}
\usepackage[dvips]{color}
\usepackage{epsfig,amssymb,amsmath}
\newcommand{\beq}{\begin{equation}}
\newcommand{\eeq}{\end{equation}}
\newcommand{\beqn}{\begin{eqnarray}}
\newcommand{\eeqn}{\end{eqnarray}}

\begin{document}
\title{Distributed MIMO Radar Using Compressive Sampling }
\author{\authorblockN{Athina P.
Petropulu,  Yao Yu and H. Vincent Poor$^+$}\\
\authorblockA{ Electrical \& Computer Engineering Department, Drexel University\\
$^+$ School of Engineering and Applied Science, Princeton
University}
\thanks{This work was supported by the Office of Naval Research under Grant
ONR-N-00014-07-1-0500.} \thanks{ This work was supported in part by
the Office of Naval Research under Grant ONR-N-00014-07-1-0500.}}
\maketitle

\begin{abstract}
A distributed MIMO radar is considered, in which the transmit and receive antennas belong to nodes of a small scale wireless network.
The transmit waveforms could be uncorrelated, or correlated in order to achieve a desirable beampattern.
The concept of compressive sampling is employed at the receive nodes
in order to perform direction of arrival (DOA) estimation.
 According to the theory of compressive sampling, a signal that is
sparse in some domain can be recovered based on far fewer samples
than required by the Nyquist sampling theorem. The DOAs of targets
form a sparse vector in the angle space, and therefore, compressive
sampling can be applied for DOA estimation. The proposed approach
achieves the superior resolution of MIMO radar with far fewer
samples than other approaches. This is particularly useful in a
distributed scenario, in which the results at each receive node need
to be transmitted to a fusion center.


\end{abstract}
\section{Introduction}

 A  multiple-input multiple-output
(MIMO) radar system, as  originally proposed in \cite{Fishler:04}-\cite{Chen:08}  transmits  multiple
independent waveforms via its antennas.
 Consider a MIMO radar  equipped with $M_t$ transmit and $M_r$
receive antennas that are close to each other relative to the
target. The phase differences induced by transmit
and receive antennas can  be exploited to form a long virtual array
with $M_tM_r$ elements. This enables the MIMO radar system to
achieve superior spatial resolution as compared  to a traditional
radar system.
 MIMO radar transmitting correlated signal waveforms  in
 order
to achieve a desired beampattern has also been proposed \cite{Stoica:07}-\cite{Fuhrmann:08}. This is useful in cases
where the radar system wishes to avoid certain directions, because
they either correspond to eavesdroppers, or are known to be of no interest.

Compressive sensing (CS) has  received considerable attention recently,
and has been applied successfully in diverse fields, e.g., image
processing \cite{Romberg:08} and wireless communications
\cite{Bajwa:06}. The theory of CS states that a $K$-sparse signal
$\mathbf{x}$ of length $N$ can be recovered exactly with high probability from
$\mathcal{O}(K\log N)$ measurements via
linear programming. Let $\Psi$ denote the basis matrix that spans
this sparse space, and let $\Phi$ denote a measurement matrix. The  convex
optimization problem arising from CS is formulated as follows:
\begin{eqnarray}
\min\|\mathbf{s}\|_1,\ \  \text{subject}\ \text{to}\ {\bf
y}=\Phi{\bf x} =\Phi \Psi {\bf s}
\end{eqnarray}
where $\mathbf{s}$ is a sparse vector with K principal elements and
the remaining elements can be ignored;  $\Phi$ is an $M\times N$
matrix incoherent with  $\Psi$ and $M\ll N$.

In this paper, we propose a distributed MIMO radar system, where  transmit and receive antennas belong to nodes of a wireless network that are uniformly distributed on a disk or a certain radius.
 located wireless network nodes.
The readings of the receive nodes are transmitted to a central node for DOA estimation.
Energy efficiency is an important issue in such a wireless network as the nodes operate on battery.
We employ the idea of compressive sampling in order to save in energy
consumed during data transmission to the central node.
Recently, the work of \cite{Gurbuz:08} considered DOA estimation of
signal sources using CS. In \cite{Gurbuz:08}, the basis matrix
$\Psi$ is formed by the discretization of the angle space. The
source signals were assumed to be unknown, and an approximate version of
the basis matrix was obtained based on the signal received by a
reference vector. The signal at the reference sensor would have to be sampled at a very
high rate  in order to construct a good basis matrix. Here, we extend the
idea of \cite{Gurbuz:08} to  the problem of  DOA estimation  for
MIMO radar.
Since the number of targets is typically
smaller than the number of snapshots that can  be obtained, DOA
estimation  can be formulated as the recovery of a sparse vector
using CS. Unlike the scenario considered in \cite{Gurbuz:08}, in
MIMO radar the transmitted waveforms are known at each  receive
antennas. This enables each receive antenna to construct the basis
matrix locally, without knowledge of the received signal at a reference sensor or any
other antenna.
 We consider the more general case of correlated signal waveforms.
We
provide analytical expressions for the average signal-to-jammer ratio (SJR)
for the proposed approach. Simulation results
show that the proposed approach can accomplish the super-resolution of
MIMO radar systems while using far fewer samples than existing methods,
such as Capon, amplitude and phase estimation (APES) and generalized
likelihood ratio test (GLRT) \cite{Xu:06}.
 In particular, the proposed approach can enable each
node to obtain a good DOA estimate independently. Further, it
results in much less information to be transmitted to a fusion center,
thus enabling savings in terms of transmission energy.

\section{Signal Model for MIMO Radar}
\begin{figure}
  \begin{center}
     \scalebox{0.25}{\includegraphics{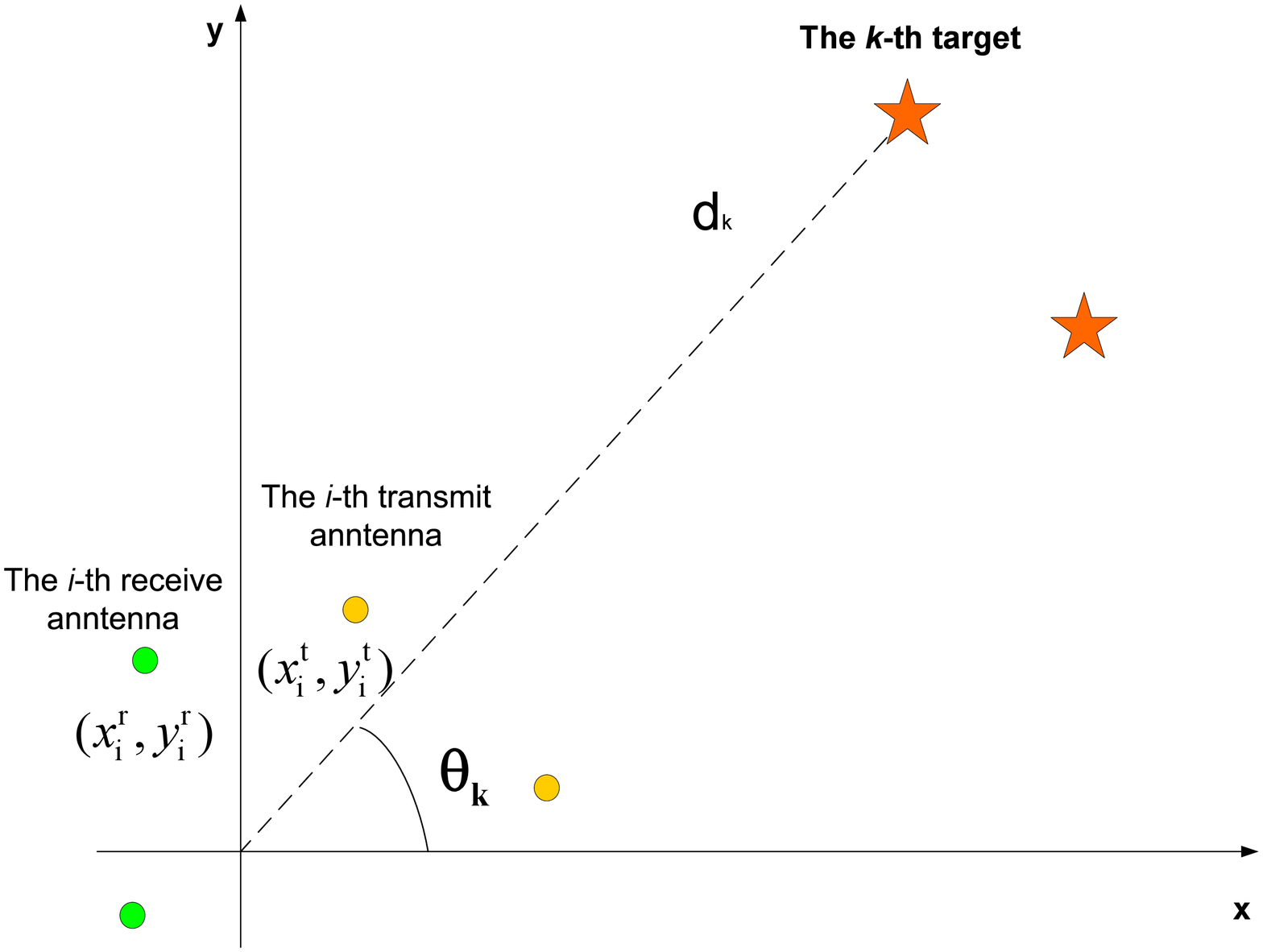}}\end{center}
 \caption{MIMO Radar System}\label{mimo_radar}
\end{figure}

We consider a MIMO radar system with $M_t$ transmit nodes and $M_r$
receive nodes that are uniformly distributed on a disk of radius
$r$.
 For simplicity,
we assume that targets and nodes lie on the same plane. Further, we
assume that each node in the network knows which are the nodes that
serve are transmit  and receive antennas and what their coordinates
are relative to a fixed point in the network. This information can
be provided by a higher network layer.  Let us denote the locations
in rectangular coordinates of the $i$-th transmit and receive
antenna by $(x^t_i, y^t_i)$ and $(x^r_i, y^r_i)$, respectively (see Fig.\ref{mimo_radar}).

The location of the $k$-th target is denoted by the polar coordinates $(d_k,\theta_k)$,
where $d_k$ is the distance between this target and  the origin,
and $\theta_k$ is the azimuthal angle, which is the unknown
parameter to be estimated in this paper.  Under the far-field
assumption
 $d_{k} \gg \sqrt{(x^t_i)^2+(y^t_i)^2}$ and $d_{k} \gg\sqrt{(x^r_i)^2+
 (y^r_i)^2}$, the distance between the $i$th transmit/receive
 antenna and the $k$-th target
 $d^t_{ik}$/$d^r_{ik}$ can be approximated as
$d^{t/r}_{ik} \approx d_k - {\eta_i^{t/r}(\theta_k)} $, where
$\eta_i^{t/r}(\theta_k)={x^{t/r}_i \cos(\theta_k)+y^{t/r}_i
\sin(\theta_k)}$.

Let $x_i(n)$ denote the discrete-time waveform transmitted by the
$i$-th transmit antenna.  Assuming the transmitted waveforms are
narrowband and the propagation is non-dispersive, the received
baseband signal at the $k$-th target equals \cite{Li:07}
\begin{eqnarray}
y_{k}(n)&=& \beta_k\sum_{i=1}^{M_t} x_i(n) e^{-j\frac{2\pi}{\lambda}
d^t_{ik}} \nonumber\\&=&\beta_ke^{-j\frac{2\pi}{\lambda}d_k}{\bf
x}^T(n){\bf v}(\theta_k)\ \ k=1,\ldots, K
\end{eqnarray}
where $\lambda$ is the transmitted signal wavelength,
\begin{eqnarray}
{\bf
v}(\theta_k)&=&[e^{j\frac{2\pi}{\lambda}\eta^t_1(\theta_k)},...,e^{j\frac{2\pi}{\lambda}\eta^t_{M_t}(\theta_k)}]^T\\
{\bf x}(n)&=&[x_1(n),...,x_{M_t}(n)]^T.
\end{eqnarray}

Due to reflection by the target, the $l$-th antenna element
receives
\begin{eqnarray}
z_{l}(n)&=&\sum_{k=1}^{K}e^{-j\frac{2\pi}{\lambda}d^r_{lk}}y_k(n)+\epsilon_{l}(n),\
l=1,\ldots, M_r
\end{eqnarray}
where $\epsilon_{l}(n)$ represents independent and identically distributed  (i.i.d.) Gaussian noise with
variance $\sigma^2$.

On letting $L$ denote the number of snapshots, we have
\begin{eqnarray}
{\bf z}_{l}&=&\left[
                \begin{array}{c}
                  z_l(0) \\
                  \vdots \\
                  z_l(L-1) \\
                \end{array}
              \right]=
\sum_{k=1}^{K}e^{-j\frac{2\pi}{\lambda}d^r_{lk}}{\bf
y}_k+{\bf e}_l\nonumber\\
&=&\sum_{k=1}^{K}e^{-j\frac{2\pi}{\lambda}(2d_k-
\eta^r_{l}(\theta_k))}\beta_k{{\bf X}}{\bf v}(\theta_k)+{\bf e}_l
\end{eqnarray}
where ${\bf y}_k=[y_k(0), \ldots, y_k(L-1)]^T$, ${\bf
e}_l=[\epsilon_{l}(0), \ldots, \epsilon_{l}(L-1)]^T$ and ${\bf X}=[
{\bf x}(0), \ldots, {\bf x}(L-1)]^T $.

By discretizing the angle space as
$\mathbf{a}=[\alpha_1,\ldots,\alpha_N]$, we can rewrite
(\ref{received signal}) as
\begin{eqnarray}\label{received signal}
{\bf z}_{l}=\sum_{n=1}^{N}e^{j\frac{2\pi}{\lambda}
\eta^r_{l}(\alpha_n)}s_n{{\bf X}}{\bf v}(\alpha_n)+{\bf e}_l
\end{eqnarray}
where
\begin{equation*}
s_n = \left\{
\begin{array}{rl}
e^{-j\frac{4\pi}{\lambda}d_{k}}\beta_k &  \text{if there is target at}\ \alpha_n \\
0  & \text{otherwise}
\end{array} \right.  \ .
\end{equation*}

\section{Compressive Sensing for MIMO Radar}
Assuming that there exists a small number of targets, the DOAs are
sparse in the angle space, i.e., $\mathbf{s}=[s_1,\ldots,s_N]$ is a
sparse vector. A non-zero element with index $j$  in ${\bf s}$
indicates that there is a target at the angle $\alpha_j$.

By CS theory,  we can construct a basis matrix $\Psi_l$ for the
$l$-th antenna  as
\begin{eqnarray}
\Psi_l=[e^{j\frac{2\pi}{\lambda} \eta^r_{l}(\alpha_1)}{{\bf X}}{\bf
v}(\alpha_1),\ldots,e^{j\frac{2\pi}{\lambda}
\eta^r_{l}(\alpha_N)}{{\bf X}}{\bf v}(\alpha_N)] \ .
\end{eqnarray}

  Ignoring the noise,  we have
$\mathbf{z}_l=\Psi_l\mathbf{s}$. Then we measure linear projections
of the received signal at the $l$-th antenna as
\begin{eqnarray} \label{receive_sig}{\bf
r}_l=\Phi_l\mathbf{z}_l={\Phi_l\Psi_l}\mathbf{s},
\end{eqnarray}
where $\Phi_l$ is an $M\times L$ random Gaussian matrix which has
small correlation with ${\Psi}_l$. Combining  the output of  $N_r$
receive antennas, we have
\begin{eqnarray}\label{cs}
{\bf r}=\left[
          \begin{array}{c}
            {\bf r}_1 \\
            \vdots \\
            {\bf r}_{N_r}\\
          \end{array}
        \right]
=\underbrace{\left[
          \begin{array}{c}
           {\Phi_1\Psi_1}  \\
            \vdots \\
           \Phi_{N_r}\Psi_{N_r}\\
          \end{array}
        \right]}_{\Theta}\mathbf{s},\ 1\leq Nr\leq M_r \ .
\end{eqnarray}
Therefore, we can recover $\mathbf{s}$ by applying the Dantzig selector
to the convex problem in (\ref{cs}) as in  \cite{Candes:07}:
\begin{eqnarray}\label{Dantzig}
\hat{{\bf s}}= \min\|{\bf s}\|_1\ \ \ s.t. \|{\Theta}^H({\bf
r}-\Theta{\bf s})\|_{\infty}<\mu.
\end{eqnarray}
According to \cite{Candes:07}, we can recover the sparse vector
${\bf s}$ with very high probability if we select
$\mu=(1+t^{-1})\sqrt{2\log N\sigma^2}$, where $t$ is a positive
scalar and  $\sigma^2$ is the noise power.

\section{Performance Analysis in the presence of a jammer signal}

In the presence of a jammer at location $(d,\theta)$ the signal received at the $l$-th
receive antenna can be represented as
\begin{eqnarray}
{\bf r}_{l}&=&\Phi_l\sum_{k=1}^{K}e^{-j\frac{2\pi}{\lambda}(2d_k-
\eta^r_{l}(\theta_k))}\beta_k{{\bf X}}{\bf
v}(\theta_k)\nonumber\\&&+\Phi_le^{-j\frac{2\pi}{\lambda}(d-
\eta^r_{l}(\theta))}\beta{{\bf b}}+\Phi_l{\bf e}_l \ .
\end{eqnarray}
where  $\beta, \ \bf b$ denote respectively the reflection amplitude
and waveform of this jammer. Since $\bf b$ is uncorrelated with the
transmitted waveforms $\bf X$, the effect of the  jammer signal is
similar to that of  addictive noise. Let ${\bf A}_l=\Phi_l^H\Phi_l$
and
 ${\bf D}_l={\bf X}^H{\bf A}_l{\bf X}$, where ${\bf
D}(i,j)$ denotes the $(i, j)$th element of ${\bf D}$. We assume that
the TX/RX nodes are uniformly distributed on a disk with the radius
$r$. Thus, the average power of the desirable signal $P_s(l)$  can
be represented by
\begin{eqnarray} P_s(l)&=&
E\{\sum_{k,k'=1}^{K}\underbrace{e^{j\frac{2\pi}{\lambda}[2(d_k-d_{k'})-(\eta^r_l(\theta_k)-\eta^r_l(\theta_{k'}))}}
_{\rho_{l}(k,{k'})}\beta_k^*\beta_{k'}\nonumber\\&&\times\underbrace{{\bf
v}^H(\theta_k) {{\bf X}}^H{\bf A}_l {\bf
X}{\bf v}(\theta_{k'})}_{Q_{kk'}}\}={E\{\sum_{k=1}^{K}|\beta_k|^2 Q_{kk}\}}\nonumber\\
&&+ E\{\sum_{k\neq k'}\rho_{l}(k,{k'})\beta_k^*\beta_{k'} Q_{kk'}\}
\end{eqnarray}
where $Q_{kk'}=\sum_{i,j}{\bf D}_l(i,j)e^{j\frac{2\pi
r}{\lambda}(\eta_j^t(\theta_{k'})-\eta_i^t(\theta_k))}$.

Following \cite{Ochiai}, we know that
$f_h(h)=\frac{2}{\pi}\sqrt{1-h^2},-1<h<1$ if
$h=\frac{\tilde{r}}{r}sin(\Psi)$, where
$f_{\tilde{r}}(\tilde{r})=\frac{2\tilde{r}}{r^2}, 0<\tilde{r}<r$ and
$f_{\psi}(\psi)=\frac{1}{2\pi}, -\pi<\psi<\pi$. Then $E\{e^{j\alpha
h}\}=2\frac{J_1(\alpha)}{\alpha}$, where $J_1(\cdot)$ is the
first-order Bessel function of the first kind. Using this property
and letting
$a_{ij}=\eta_j^{t/r}(\theta_{k'})-\eta_i^{t/r}(\theta_k)$, we have
 \begin{eqnarray}
E\{e^{j\frac{2\pi r}{\lambda}a_{ij}}\}= \left\{
\begin{array}{rl}
\eta(4\sin(\frac{\theta_{k'}-\theta_k}{2}))& i=j\\
\eta^2(2) &  i\neq j
\end{array} \right.
 \end{eqnarray}
where $\eta(x)=2\frac{J_1(x\frac{\pi r}{\lambda})}{x\frac{\pi
r}{\lambda}}$.

Therefore, the average power of the desirable signal $P_s(l)$ taken
over the positions of TX/RX nodes can be found to be:
\begin{eqnarray}\label{signal_power1}
P_s(l)&=&\sum_{k=1}^K |\beta_k|^2 [\sum_{i}{\bf D}_l(i,i)+
\sum_{i\neq j}{\bf
D}_l(i,j)\eta^2(2)]\nonumber\\
&+&\sum_{k\neq k'}\beta_k^*\beta_{k'}e^{j\frac{4\pi}{\lambda}(d_k-d_{k'})}\underbrace{\eta(4\sin(\frac{\theta_{k'}-\theta_k}{2}))}_{\eta_{kk'}}\nonumber\\
&\times&[\eta_{kk'}\sum_{i}{\bf D}_l(i,i)+ \sum_{i\neq j}{\bf
D}_l(i,j)\eta^2(2)].
\end{eqnarray}

Similarly, the power of the jammer signal is given by
\begin{eqnarray}\label{jammer_power}
P_j(l)&=& (e^{-j\frac{2\pi}{\lambda}(d-\eta^r_l(\theta))}
\beta)(e^{-j\frac{2\pi}{\lambda}(d-\eta^r_l(\theta))} \beta)^*\nonumber\\
&&\times{\bf b}^H{\bf A}_l{\bf b}=|\beta|^2{\bf b}^H{\bf A}_l{\bf b}
\ .
\end{eqnarray}

The SJR given the node locations is the ratio of the power of the
signal over the power of the jammer. Since the denominator does not
depend on node locations, the average SJR equals the ratio of
(\ref{signal_power1}) and (\ref{jammer_power}).

Since the jammer signal is uncorrelated with the transmitted signal,
the SJR can be improved by correlating the jammer signal with the
transmitted signal. Combining this with CS, the measurement matrix
in (\ref{receive_sig}) is modified as 
\begin{eqnarray}\label{measurement_matrix}
\tilde{\Phi}_l=\Phi_l{\bf X}^H \ .
\end{eqnarray}
Moreover, since  $\Phi_l$ is a Gaussian random matrix, $\tilde{\Phi}_l$ is
still Gaussian; therefore it satisfies the restricted isometry
property (RIP)  and is incoherent with
 $\Psi_l$, thus guaranteeing a stable solution to
(\ref{Dantzig}). Based on (\ref{measurement_matrix}),
the average power of the desirable signal $P_s(l)$ is equal to
(\ref{signal_power1}) except ${\bf D}_l={\bf X}^H{\bf XA}_l{\bf
X}^H{\bf X}$. The average power of the jammer signal using
$\tilde{\Phi}_l$ is rewritten as  $P_j(l)= |\beta|^2{\bf b}^H{\bf
X}{\bf A}_l{\bf X}^H{\bf b}$.

Approximating ${\bf X}^H{\bf X}\sim {\bf I}_{M_t}$ and using ${\bf
b}^H{\bf b}=1$, the SJRs based on $\Phi_l$ and $\tilde{\Phi}_l$ can be
approximated as $\frac{M_t\sum_{k=1}^K |\beta_k|^2}{|\beta|^2} $ and
$\frac{L\sum_{k=1}^K |\beta_k|^2}{|\beta|^2}$, respectively.
Therefore, the SJR using (\ref{measurement_matrix}) can be generally
improved by a factor of $L/M_t$
 since $L\gg M_t$.  the DOA estimates can be improved by the increase in $L$.  However,
the time duration of the radar pulse might need to be longer as
well.

As simulation results show (see Section \ref{simulation}),
the proposed method can yield  good performance even using a single
receive antenna. With a good initial estimate of DOA, the receive
nodes can adaptively refine their estimates by constructing  a
higher resolution basis matrix $\Psi_l$ around that DOA.
 Restricting the candidate angle space,
 may reduce the samples in the angle space that are required
for constructing the basis matrix, thus reducing the complexity of
the $\ell_1$ minimization step. On the other hand, the resolution of
target detection can be improved by taking the denser samples of the
angle space around the intimal DOA estimate. Furthermore, the
transmit node can design the correlated waveforms for transmit
beamforming as well based on the good initial estimate.

\section{Simulation Results }\label{simulation}

In this section, we consider a MIMO radar system with the
transmit/receive antennas uniformly distributed on a disk of radius
$10$m.  The number of transmit nodes is fixed at $M_t=50$. The
carrier frequency is 8.62 GHz.
 A maximum of $L = 512$ snapshots are
considered at the receive node. The received signal is
 corrupted by  zero mean Gaussian noise.
 The SNR is
set to 20 dB .

 There are two targets located at
$\theta_k=-1\textordmasculine,\ 1\textordmasculine$, with reflection
coefficients $\beta_k=1, k=1,2$. A jammer is located at
$15\textordmasculine$ and transmits an unknown Gaussian random waveform and
with amplitude 20, i.e., 26 dB above the target reflection
coefficients
 $\beta_k$. We sample the angle space by increments of $0.5\textordmasculine$
from $-8\textordmasculine$ to $8\textordmasculine$, i.e., ${\bf
a}=[-8\textordmasculine,-7.5\textordmasculine,\ldots,7.5\textordmasculine,8\textordmasculine]$.

\begin{figure}
  \begin{center}
    \scalebox{0.28}{\includegraphics{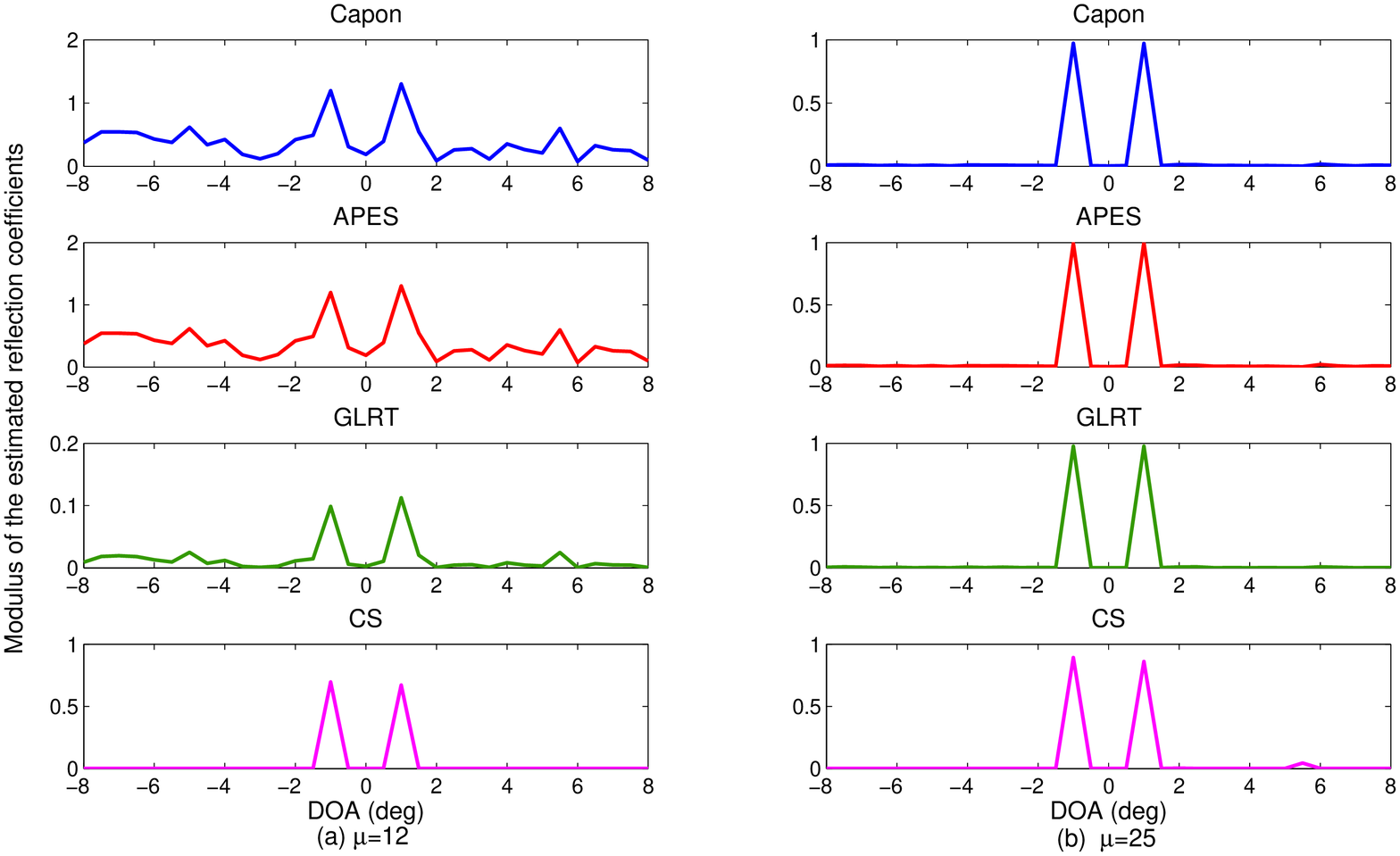}}\end{center}
 \caption{DOA estimates of two targets with $1$ (left column) and $10$ (right column) receive antenna using independent waveforms. The top three curves were obtained using $512$ snapshots.
 The bottom curve was obtained using $35$ snapshots only.
  }\label{rec_unco}
\end{figure}

\begin{figure}
  \begin{center}
      \scalebox{0.28}{\includegraphics{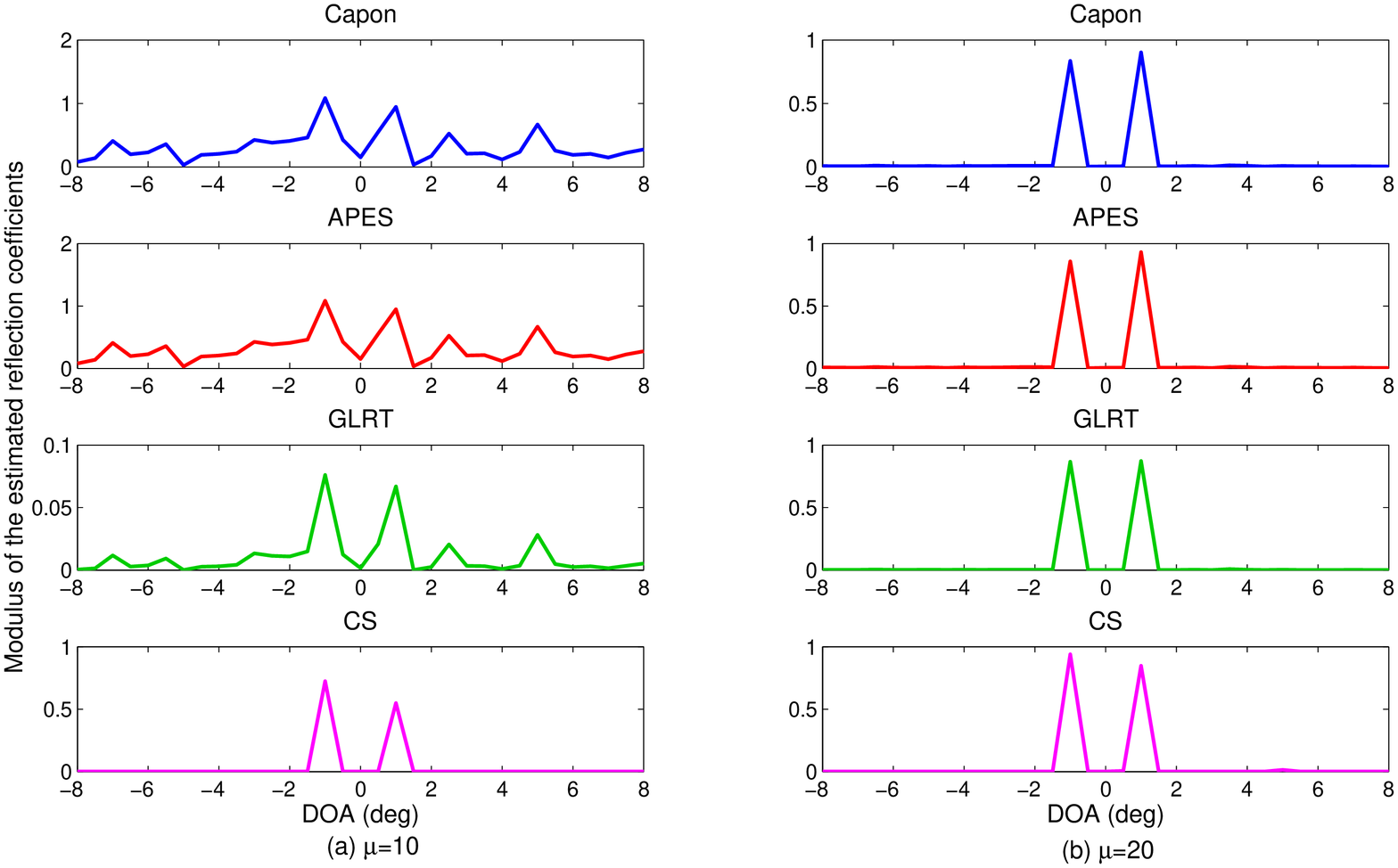}}\end{center}
 \caption{DOA estimates of two targets with close azimuthal angles using $1$ (left column) and $10$ (right column) receive antenna using independent waveforms. The top three curves were obtained using $512$ snapshots.
 The bottom curve was obtained using $35$ snapshots only. }\label{rec_co}
\end{figure}

 First, we compare the performance of DOA estimation using
the proposed method and three approaches \cite{Xu:06}, i.e., the
Capon, APES and GLRT techniques. Fig. \ref{rec_unco} and Fig.
\ref{rec_co} show the modulii of the estimated reflection
coefficients $\beta_k$, as functions of the azimuthal angle for
$N_r=1$ and $10$ receive antennas, respectively. In Fig.
\ref{rec_unco}, we use the uncorrelated QPSK waveforms; while in
Fig. \ref{rec_co}, we use correlated waveforms designed according to
the desired beampattern $P_d(\alpha_n)$ as
 \begin{eqnarray}\label{beam_des}
P_d(\alpha_n)= \left\{
\begin{array}{rl}
1 & -3\textordmasculine \leq \alpha_n\leq 3\textordmasculine\\
0  &  -8\textordmasculine\leq \alpha_n < -3\textordmasculine \
\text{and}\  3\textordmasculine< \alpha_n \leq 8\textordmasculine
\end{array} \right.  \ .
 \end{eqnarray}
Based on that beampattern, the method of \cite{Stoica:07}  was
followed to design ${\bf R}$. Then the transmitted waveforms  can be
constructed  as ${\bf x(n)}={\bf R}^{\frac{1}{2}}{\bf w}$, where
${\bf w}$ is a $\text{i.i.d}$ random vector with zero mean and
$E\{{\bf w}{\bf w}^H\}={\bf I}/L$.

 In both (a) and (b), the top
three curves correspond to the azimuthal estimates obtained via
Capon, APES and GLRT, using $512$ snapshots. The bottom curve is the
result of the proposed approach, obtained using $35$ snapshots only.
One can  see that in the case of using only one receive node, the
presence of the two targets is clearly evident via the proposed
method based on $35$ snapshots only using both independent and
correlated waveforms. The other methods produce spurious peaks away
from the target locations. When the measurements of multiple receive
nodes are used at a fusion center, the proposed approach can yield
similar performance to the other three methods. However, the
comparison methods would have to transmit to the fusion center $512$
received samples each, while in the proposed approach, each node
would need to transmit $35$ samples each.

The threshold $\mu$ in (\ref{Dantzig})  affects DOA estimation for
the proposed method. The increase in $\mu$  while
keeping $M_t$ and $N_t$ constant  can reduce the ripples of DOA
estimates at the non-target azimuth angles at the expense of
 the accuracy of the target-reflection-coefficient
estimates.  The increase in $\mu$ can also reduce the complexity of
(\ref{Dantzig}) because the constraint is  looser than that of
smaller $\mu$. If $\mu$ is too large, however, the $\ell_1$-norm
minimization does not work. In Fig.\ref{rec_unco} and \ref{rec_co},
 relatively large thresholds, i.e.,  $\mu=12,10$, were used for the
single receive node case. As a result, the CS method yielded less
accurate estimates of the reflection coefficients magnitude
 than the Capon and APES, but with
 very few ripples.

\begin{figure}
  \begin{center}
     \scalebox{0.30}{\includegraphics{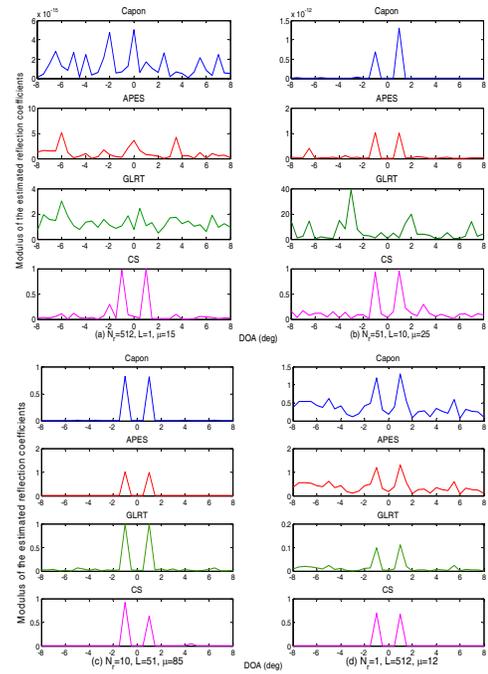}}\end{center}
 \caption{DOA estimates with different sets of $N_r$ and $L$ }\label{com_N_r_L}
\end{figure}

\begin{figure}
  \begin{center}
     \scalebox{0.28}{\includegraphics{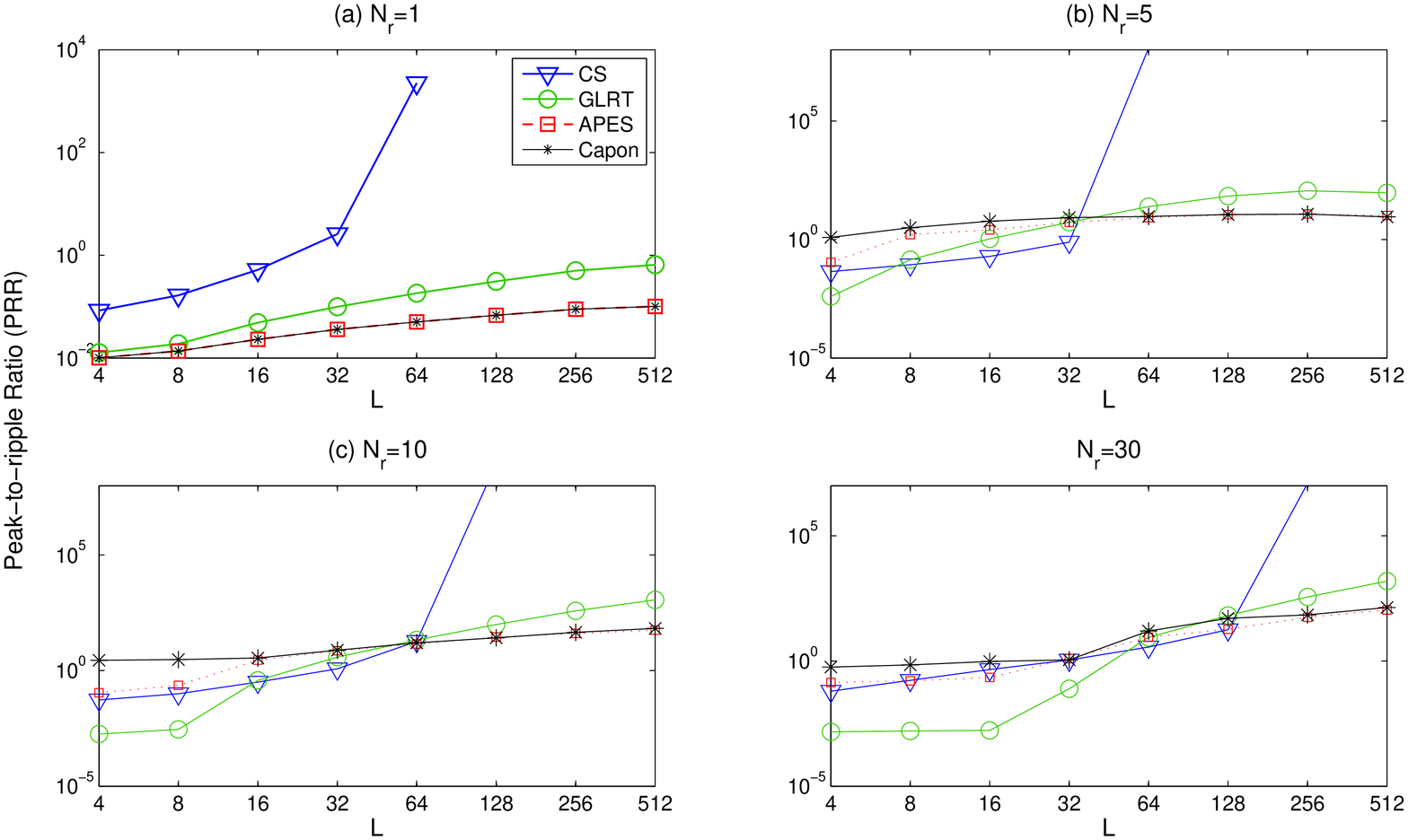}}\end{center}
 \caption{Peak-to-ripple Ratio vs. $L$ with different $N_r$  }\label{com_peak_ripple_ratio}
\end{figure}

Finally, we discuss the effect of $L$, $N_t$ and $M_t$  on the
performance of the Capon, APES, GLRT and CS. Fig. \ref{com_N_r_L}
compares the performance of these four approaches using
independent waveforms for different combinations of $N_r$ and $ L$,
whose product is fixed at $512$.  In order to quantify the
performance of DOA estimation, we define the ratio of the square
amplitude  of the DOA estimate at the  target azimuth angle to the
sum of the square amplitude of DOA estimates at other angles as the
peak-to-ripple ratio (PRR). Fig. \ref{com_peak_ripple_ratio}
compares PRR  as a function of $L$ for  these four approaches using
uncorrelated signal waveforms. We consider the scenarios in which
$N_r=1,5,10,30$. For fixed $N_t$ and $M_t$, the increase in $L$ can
improve the performance of these four methods. In the presence of a
moderate jammer, APES and CS can yield  relatively accurate DOA
estimates even with a small $L$.
 For Capon and
GLRT, $N_r$ must be greater than $L$ in order to obtain a
nonsingular sample covariance matrix  of the received signal
$\tilde{\bf R}$. This is because Capon and GLRT need to calculate
the inverse of $\tilde{\bf R}$.  On the other hand, an increase
 in both  $M_t$ and $N_r$ can also improve the performance of the
Capon, APES, GLRT and CS while $L$ is fixed. If either  $M_t$ or
$N_r$ is too small, even a significant increase in the other
parameter cannot improve performance of the first three approaches.
However, CS can yield the desired DOA estimates even with a single
receive antenna with a sufficient $L$ and $N_t$. In the senecios
considered in our simulations where $L=2^9$, for instance, $M_t$ and
$N_r$ are required to  be greater than $8$ to yield the desired DOA
estimates using the Capon, APES and GLRT. With a single receive
node, CS requires at least $20$ transmit nodes, while in the cases
of multiple receive nodes, the requirement of $M_t$ and $N_r$ for CS
is the same as in the other three methods.

\section{Conclusion}
We proposed a distributed MIMO Radar system implemented by  a small
size randomly dispersed wireless network. There are several advantages in using
the proposed distributed approach as opposed to using a standard
linear array. The radar system can be easily deployed;  no pre-existing infrastructure is required.
In a high density network there are many degrees of freedom to design the beampattern as desired around the look direction, which is important for clutter reduction or for reduction of scanning time. By randomizing the set of transmitters and receivers we can used the network power efficiently. By selecting well separated nodes we can increase spatial diversity. The resolution can be easily adjusted by employing more or less transmit nodes. The radar system is robust; should some nodes be deactivated the system performance will not be affected.

For the proposed MIMO
radar system,  a compressive sensing method has been exploited to
estimate the DOAs of targets  using both independent and  correlated
waveforms. The DOA of targets can construct a sparse vector in the
angle space. Therefore, we can solve for this sparse vector by
$\ell_l$-norm minimization with many fewer samples than conventional
methods, i.e. the Capon, APES and GLRT techniques. The proposed
method  is superior to these conventional methods when one receive
antenna is active. If multiple receive antennas are used, the
proposed approach can yield similar performance  to the other three
methods, but by using far fewer samples.

\bigskip
\centerline{\bf Acknowledgment} The authors would like to thank Dr.
Rabinder Madan of the Office of Naval Research for bringing the possibility of using compressive
sampling for angle-of-arrival estimation to their attention.

\bibliographystyle{IEEE}

\end{document}